 \documentclass[final,5p,times,twocolumn,numbers]{elsarticle}

\usepackage{amssymb}
\usepackage{lipsum}
\usepackage{amsmath}
\usepackage{graphicx} 
\usepackage{lipsum}
\usepackage{booktabs}
\usepackage{enumitem}
\usepackage{tabularx}

\makeatletter
\def\@preprintinfo{}

\def\ps@firstpage{%
  \let\@oddhead\@empty
  \let\@evenhead\@empty%
  \def\@oddfoot{\hfill\thepage\quad\the\year\hfill}%
  \def\@evenfoot{\hfill\thepage\quad\the\year\hfill}%
}

\AtBeginDocument{\thispagestyle{firstpage}}
\makeatother

\begin{document}

\begin{frontmatter}

\title{Modeling Epidemic Dynamics of Mutant Strains with Evolutionary Game-based Vaccination Behavior }


\author[1]{Wenjie Zhang}
\author[2]{Yusheng Li  }
\author[3]{Qin Li \corref{cor1}}
\author[2]{Guojun Huang\corref{cor1}}
\author[2]{Minyu Feng }
\cortext[cor1]{Corresponding authors. E-mail address: qinli1022@swu.edu.cn, huangguojun@swu.edu.cn}

\affiliation[1]{organization={Innovation and Entrepreneurship College (Han Hong College)},
    addressline={Southwest University}, 
    city={Chongqing},
    postcode={400715}, 
    country={China}}
\affiliation[2]{organization={ College of Artificial Intelligence},
    addressline={Southwest University}, 
    city={Chongqing},
    postcode={400715}, 
    country={China}}
\affiliation[3]{organization={Business College},
    addressline={Southwest University}, 
    city={Chongqing},
    postcode={402460}, 
    country={China}}            

\begin{abstract}
The outbreak of mutant strains and vaccination behaviors have been the focus of recent epidemiological research, but most existing epidemic models failed to simultaneously capture viral mutation and consider the complexity and behavioral dynamics of vaccination. To address this gap, we develop an extended SIRS model that distinguishes infections with the original strain and a mutant strain, and explicitly introduces a vaccinated compartment state. At the behavioral level, we employ evolutionary game theory to model individual vaccination decisions, where strategies are determined by both neighbors' choices and the current epidemiological situation. This process corresponds to the time-varying vaccination rate of susceptible individuals transitioning to vaccinated individuals at the epidemic spreading level. We then couple the epidemic and vaccination behavioral processes through the microscopic Markov chain approach (MMCA) and finally investigate the evolutionary dynamics via numerical simulations. The results show that our framework can effectively mitigate outbreaks across different disease scenarios. Sensitivity analysis further reveals that vaccination uptake is most strongly influenced by vaccine cost, efficacy, and perceived risk of side effects. Overall, this behavior-aware modeling framework captures the co-evolution of viral mutation and vaccination behavior, providing quantitative and theoretical support for designing effective public health vaccination policies.
\end{abstract}



\begin{keyword}
Epidemic dynamics \sep  Evolutionary games\sep Vaccination behavior\sep Mutant Strains


\end{keyword}

\end{frontmatter}




\section{Introduction}
\label{introduction}
In human history, epidemic prevention and control have always been a critical and challenging issue globally. Major epidemics throughout history, such as the Black Death, Severe Acute Respiratory Syndrome (SARS), and the recent Coronavirus Disease 2019 (COVID-19) pandemic, have had profound impacts on human society both materially and psychologically \cite{cohen2003efficient}. Epidemics have, to some extent, influenced the course of history. Under such severe circumstances, accurately modeling the spread of epidemics has become a core research direction for formulating effective intervention strategies. Especially as vaccination behaviors become increasingly complex \cite{cai2014effect, yang2022impact, wang2017vaccination}, and the continuous emergence of mutants such as Omicron \cite{di2018multiple}, traditional modeling methods face tremendous challenges, highlighting the urgent need in academia for more refined and realistic models.

For a long time, compartmental models have been the foundation of epidemic modeling, especially since the classical SIR (Susceptible-Infectious-Recovered) model and SIS (Susceptible-Infectious-Susceptible) model proposed by Kermack and McKendrick \cite{kermack1927contribution} laid the groundwork for numerous subsequent studies, inspiring a large number of related research works \cite{liu2014information}. As research progressed \cite{jusup2022social} and more complex problems emerged, these models were extended into more refined frameworks. For example, the SEIR (Susceptible-Exposed-Infectious-Recovered) model \cite{he2020seir} and the SIRS (Susceptible-Infectious-Recovered-Susceptible) model successfully cover different stages of disease progression and changes in immunity dynamics; in addition, there are extended forms such as the SAIR (Susceptible-Asymptomatic Infection-Infectious-Recovered) model \cite{grunnill2018exploration}. Furthermore, some researchers have incorporated novel and detailed settings such as protection levels \cite{li2021protection}, further enhancing the descriptive capabilities of traditional models. Moreover, the continued development of infectious disease models has even attracted numerous researchers to adopt more innovative methods for theoretical derivation \cite{alshahrani2024reliable}.

Meanwhile, the introduction of complex network theory has further driven developments in this field. The simplicity and comprehensiveness of complex networks \cite{ji2024focus, ji2023signal} enable researchers to simulate population structures in a more intuitive, concise, and high-fidelity manner, while also fully considering the heterogeneity of people's contact patterns in real life \cite{chen2024siqrs, li2025epidemic, liu2023epidemic}. At the same time, pioneering models such as the Watts-Strogatz small-world (WS) network \cite{watts1998collective} and the Barabási-Albert scale-free (BA) network \cite{albert2002statistical} have provided unprecedented possibilities for capturing social connections. Furthermore, research on the robustness and resilience of complex networks \cite{artime2024robustness} has greatly enhanced its potential and feasibility as a foundational model for epidemic modeling. In recent years, research on epidemic spreading on complex networks has been continuously innovating, with a large number of novel models and theories emerging, including multilayer frameworks that incorporate information dissemination and cognitive effects, such as Unaware - Aware - Unaware - Susceptible - Infected - Susceptible (UAU-SIS) and Unaware - Aware - Unaware - Susceptible - Infected - Rcovered (UAU-SIR) \cite{granell2013dynamical, zheng2018interplay}, among others. There are many other advanced achievements \cite{sun2022diffusion, fan2022epidemics, boccaletti2014structure, li2021dynamics} which have greatly promoted the development of more efficient epidemic models on complex networks. 

In addition, we note that there is also a substantial amount of research attempting to integrate evolutionary game theory methods, which play a significant role in social activities and welfare \cite{han2026cooperation}, into epidemic models, opening new avenues for studying vaccination behavior \cite{wang2016statistical} and policy-making. In this field, existing studies have incorporated strategic decision-making into models, where individuals decide whether to get vaccinated based on perceived risks, costs, and social influences \cite{zanette2002effects, starnini2013immunization, dinleyici2021vaccines}. At the same time, these related models indicate that vaccination decisions are often influenced by neighbors' choices and the overall epidemic situation, thereby bridging the gap between individual-level behavior and population-level disease dynamics \cite{li2023global, tchoumi2022dynamic}. They demonstrate more precise fitting and, by leveraging the payoff-driven nature of evolutionary games, can more effectively capture the high complexity of various behaviors observed in real life.

However, existing models, despite demonstrating good fitting performance and practical value, still have several important limitations. Firstly, many studies failed to adequately differentiate between vaccinated individuals (V), susceptible individuals (S), and recovered individuals (R), often grouping the three into a single behavioral category \cite{zanette2002effects}. This simplification may confound vaccine-induced immunity with natural or temporary immunity, and cannot consider the potential role of individuals who have been vaccinated but have not yet been infected in disease transmission. In addition, although some models incorporate vaccination behavior \cite{li2023global, tchoumi2022dynamic}, they often lack a mechanistic description of individual decision-making processes. At the same time, while multi-disease models do exist, most focus on co-infection \cite{hebert2015complex}, competition \cite{karrer2011competing, yang2019competitive}, cross-immunity \cite{yang2016dynamics}, or neutral mutation prediction \cite{gubar2013optimal}, rather than treating variations as a state transition mechanism reflecting a ``parent-child'' strain relationship, as observed in real variants such as Omicron \cite{sun2022origin}. Moreover, although previous studies have attempted to integrate viral mutations within the SIR framework \cite{marquioni2021modeling}, there is still a lack of comprehensive models that simultaneously incorporate both mutation dynamics and evolutionary game-theory-based vaccination strategies.

To address these gaps, this paper makes the following contributions:

\begin{enumerate}[label=\arabic*),topsep=0pt]
    \item We extend the traditional SIRS model by introducing a vaccinated state (V) and an infected state for mutant strains ($I_2$), and integrate evolutionary game theory to construct a vaccine decision-making model, thereby simultaneously capturing variant-driven disease evolution and the behavioral adaptation process of individuals based on the payoffs from neighborhood interactions.

    \item We propose a detailed evolutionary game theory vaccination model on complex networks: in this model, individuals and their neighbors jointly act as strategy participants, evaluating the profits of vaccination based on vaccine efficacy, cost, and infection risk. The core innovation lies in constructing a vaccination update mechanism that integrates game outcomes among local neighbors with the overall epidemic state---the overall epidemic influence includes factors such as the herd effect brought by vaccinated individuals (V) and risk perceptions of individuals regarding potential infection from original strain infectors ($I_1$), variant strain infectors ($I_2$), and mutation rate ($\mu$). This two-layer mechanism allows individual decisions to be influenced by both neighborhood interactions and population-level factors (such as infection density and vaccination coverage), thereby triggering self-protection and herd effects. These mechanisms dynamically interact, achieving adaptive decision-making and realistic vaccination behavior in the process of ``evolutionary epidemic-game'' coupling.

    \item We adopt the MMCA coupling approach, integrating vaccination game theory with epidemic transmission models and related mechanisms, to simulate the state transition process and subsequently propose state transition dynamical equations. We organize model parameters and strategies to conduct extensive simulation experiments, and the simulation results show that higher risk perception and stronger herd mentality can significantly increase vaccination willingness and curb transmission. The model is effective in controlling outbreaks across multiple scenarios---diseases with low to moderate infection rates and mutation rates cannot sustain transmission, and even large-scale outbreaks can be quickly contained. These findings provide actionable recommendations for public health policy, indicating that reducing vaccine costs, improving vaccine efficacy, and alleviating adverse effects can significantly increase vaccination coverage and enhance epidemic control.
\end{enumerate}

The remainder of this paper is organized as follows: Sec. \uppercase\expandafter{\romannumeral2} introduces the structure of our epidemic model, including epidemic transmission and vaccination strategy. Then, integrate them through the MMCA. Sec. \uppercase\expandafter{\romannumeral3} presents simulation results validating the model performance and sensitivity analyses. Finally, Sec. \uppercase\expandafter{\romannumeral4} concludes the paper and discusses future research directions.

\section{Model Description}
Existing epidemic models often fail to adequately capture the interaction between vaccination behavior and viral mutation dynamics \cite{cai2012two, wang2016computational}. Many approaches either simplify the complex decision-making process of individuals regarding vaccination or overlook the influence of both individual and population levels---that is, the micro and macro aspects---on vaccination behavior. To fill these gaps, we propose a novel compartmental framework that combines evolutionary game theory for vaccination decision-making with a transmission model incorporating dual-strain mutation to study how strategic behavior affects the spread of new variants and how it is influenced by the emergence of new variants, providing a more realistic description of complex epidemic systems.
\subsection{Epidemic Spreading Model with Vaccination and Mutation}
In this subsection, we present an extended epidemic model that incorporates a vaccination state, V, and a mutant strain infection state, $I_2$, which differs from the initial infection state, $I_1$, thereby building upon the traditional SIRS framework. Drawing on insights from evolutionary game theory and epidemic spreading over complex networks, we regard each node as an individual, and edges represent interactions through which disease transmission may occur. 

To clearly distinguish between vaccination strategy and epidemic status, we define two random variables for each node i: $X_i(t)$ indicates whether node $i$ is vaccinated at the moment $t$ (1 is vaccinated and 0 is unvaccinated), and $Y_i(t)\in\{S, V, I_1, I_2, R\}$ represents the epidemic status. This distinction avoids confusion between the two types of dynamic processes. We assume that the network is a closed and undirected weighted network, and do not consider demographic processes such as birth and death for the sake of simplifying the model. Vaccination behavior occurs at the same time as the spread of the epidemic: susceptible individuals can choose to be vaccinated at any time, at which point $X_i(t)=1$.

Now, we give the definitions of the five categories of epidemic state space:
\begin{enumerate}
    \item S (Susceptible State): Uninfected and unvaccinated individuals. Such individuals may be infected by the original strain $I_1$ or the mutant strain $I_2$.
    \item V (Vaccination State): Vaccinated individuals. Although it is still possible to be infected by both strains, the vaccine will reduce the probability of infection to $\lambda$.
    \item $I_1$ (Original Strain Infection State): Individuals infected with wild-type virus. Infection has a probability $\mu$ of mutating into a mutant strain of $I_2$, and it will recover at a rate of $\gamma_1$ as well.
    \item $I_2$ (Mutant Strain Infection State): Individuals infected with the mutant strain have a recovery rate of $\gamma_2$.
    \item  R (Recovered State): Individuals who have recovered after infection. Over time, such individuals may lose their immunity and return to the S state based on the immunity loss rate $\alpha$.
\end{enumerate}
It should be noted that the immune system of individuals in R state will turn into S state after the decline of immunity for the following reasons: First, in this model, there is an essential difference between the mechanism of vaccine-induced immunity and natural immunity after recovery \cite{assis2021distinct}; Second, the immunity produced by vaccines will weaken over time and eventually disappear \cite{feikin2022duration}; Third, these two forms of immunity do not act on each other and do not overlap with each other.

To present our model more clearly and easily, we have listed the relevant symbolic explanations in Tab. 1.
\begin{table}[h]
\centering
\caption{Symbolic Explanations of The Transition Probabilities.}
\vspace{0.5em}
\begin{tabularx}{\linewidth}{l X} 
\toprule  
\textbf{Symbol} & \ \ \ \ \ \ \ \ \ \textbf{Explanation} \\
\midrule  
$\beta_1$ & Infection rate of naturally contracting disease $I_1$ \\
$\beta_2$ & Infection rate of naturally contracting disease $I_2$ \\
$\mu$  & The mutation rate of $I_1$ \\
$\theta$  & The probability of $S$ population getting vaccinated \\
$\lambda$ & The parameter of the infection probability reduced by vaccination \\
$\gamma_1$   & The recovery rate after being infected with $I_1$ \\
$\gamma_2$   & The recovery rate after being infected with $I_2$ \\
$\alpha$  & Rate of immunity loss \\
\bottomrule  
\end{tabularx}

\label{tab:symbol_three_line}
\end{table}

We can give a more direct definition of the probability of a node $i$ transitioning from state S to state V at time $t$, $\theta_i(t)$, through a conditional probability as
\begin{equation}
   \begin{split}
   \theta_i (t) = P[Y_i (t+1)=V \mid Y_i (t)=S]
   \end{split},
\end{equation}
and the specific transition mechanisms and probability expressions will be elaborated on in the next subsection. To illustrate more clearly, the state of each node evolves continuously through the coupled dynamics of epidemic propagation and vaccination game interactions, and the schematic diagram of the model is shown in Fig. 1.
\begin{figure}[ht]
    \centering
    \includegraphics[width=\linewidth]{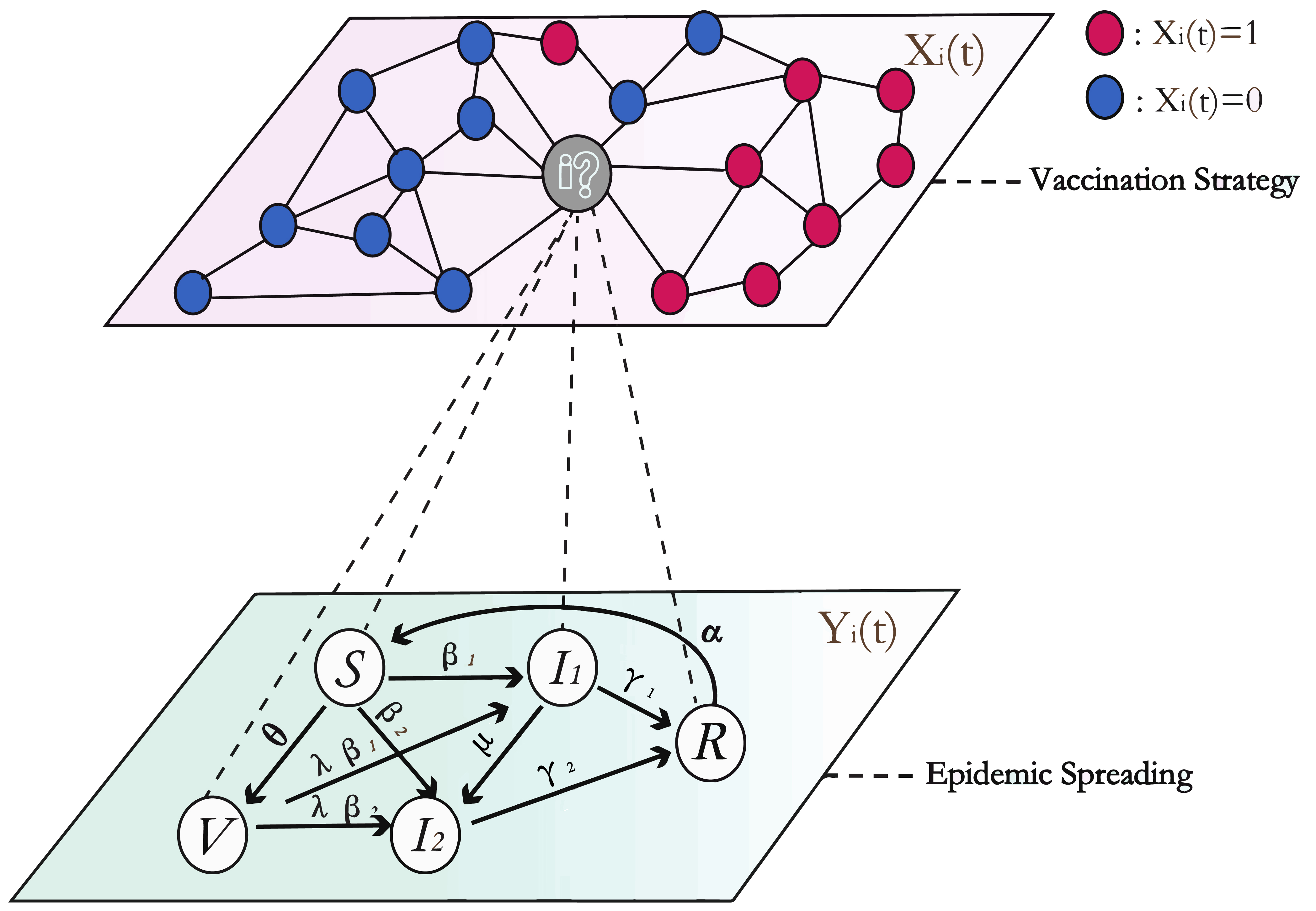}
    \caption{\textbf{Coupled dynamics of strategic vaccination and expanded SIRS epidemic spreading. }This figure illustrates the core mechanisms of strategic vaccination (above layer) and epidemic spread (below layer) and their relationship. The above subfigure illustrates a strategy update for a current individual (gray node $i$). Its decision is influenced by the two choices of its surrounding neighbors (connected to it by solid lines)---to get vaccinated (red nodes) or not (blue nodes).  Below (Disease Layer): Shows the state transitions during epidemic spreading. Solid arrows represent state transitions governed by rates ($\beta_1$, $\beta_2$, $\mu_1$, $\mu_2$) and vaccination effect ($\lambda$). The key coupling of these two layers is that the vaccination decision from the game layer decides the process of an individual in state S transitioning to the V state, then influences the whole spreading process  (demonstrated via the dashed arrows).  This figure is used to display how vaccination behavior co-evolves with the epidemic, as individuals' strategies directly alter their disease state probabilities.}
    \label{fig: enter-label}
\end{figure}

\subsection{Evolutionary Game Model for Vaccination}
To more realistically simulate the complex dynamics of vaccination decision-making during an epidemic, we construct an individual vaccination behavior model based on evolutionary game theory on complex networks. In the model, each individual and their neighbors are considered participants in a strategic game. Individuals make decisions by comparing the payoffs of ``getting vaccinated'' versus ``not getting vaccinated'', taking into account factors such as vaccine efficacy, related costs, and infection risk. Additionally, individual decisions are influenced by the overall epidemic situation: the current infection density encourages individuals to adopt self-protective behavior, while the overall vaccination coverage rate generates a herd effect. These mechanisms interact dynamically, collectively driving the evolution of vaccination probability over time.
\begin{figure*}[h]
    \centering
    \includegraphics[width=1\linewidth]{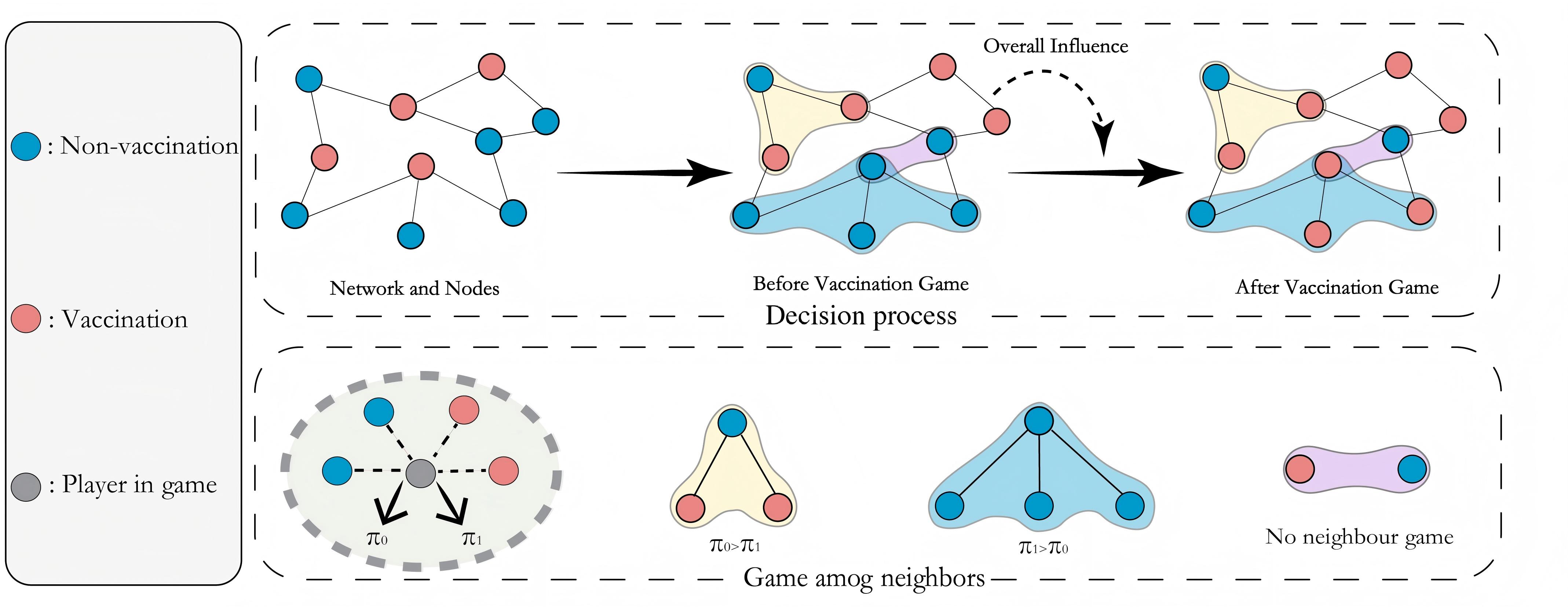}
    \caption{\textbf{Mechanism of the vaccination game:} 
This schematic elucidates the process of individual vaccination choices in our model and the details of the vaccination game among neighbors. The top row (Decision process) illustrates the population-level evolution from left to right: the initial distribution of vaccinated (red nodes) and unvaccinated (blue nodes) individuals; their intermediate states after being influenced by the epidemic dynamics before the vaccination game; and the configuration after the vaccination game. Each individual's decision is determined by both the overall epidemic influence (indicated by dashed arrows) and the outcomes of its neighbor games (detailed in the bottom row). Bottom row (Game among neighbors) tells the mechanism of the game among neighbors based on the three distinct neighbor environments (yellow, blue, and purple sets) from above. It shows that for a focal decision-maker (gray node), the vaccination strategy is analyzed for the vaccination choices of neighbors that yield distinct fitness values for it, which then inform its decision based on the corresponding payoff scenarios.}
    \label{fig:enter-label}
\end{figure*}
Our model couples the physical process of epidemic spread with the individual decision-making process regarding vaccination, enabling continuous interaction between the two. Therefore, in this subsection, we first clarify the profits of the two vaccination strategies for individuals, which will start at the individual level, analyzing the methods for calculating vaccination-related profits and the rules for strategy updating, and then exploring how these decisions affect the progress of the epidemic, specifically the transition of the individual S to the individual V. For each individual, there are only two strategies: vaccination (V) and non-vaccination (NV). We begin by defining the individual payoff matrix $\Pi$ as follows:
\begin{equation}
\begin{array}{l}
    \scriptstyle V \\ 
    \scriptstyle N    
\end{array}
\overset{
    \scriptstyle 
    \begin{array}{cc}
        \scriptstyle V & \scriptstyle N 
    \end{array}
}{
    \left(
    \begin{array}{cc}
        p_{vv} & p_{vn} \\
        p_{nv} & p_{nn}
    \end{array}
    \right)
},
\end{equation}

The specific formulas of each element are defined as follows:
\begin{equation}
\begin{cases}
   p_{vv}=w_{Id} (1-\lambda^2)-C_{V}-w_{pd}, & \\
    p_{vn}=w_{Id} (1-\lambda)-C_{V}-w_{pd}, & \\
    p_{nv}=\xi w_{Id} (1-\lambda), \\
    p_{nn}=0.
\end{cases}
\label{eq: piecewise_function}
\end{equation}
Among them, the vaccination cost $C_V$ includes direct expenses such as vaccine procurement and vaccination services. $w_{pd}$ represents the potential disutility, including possible side effects from vaccination or potential adverse effects of an individual as a result of other vaccinations considered. At the same time, $w_{Id}$ reflects the overall perceived infection risk of individuals in the context of the epidemic, which is composed of the respective perceived risks of the two major disease strains, namely, $w_{Id}=w^{I_1}_{Id}+ w^{I_2}_{Id}$. Vaccine efficacy is expressed in $\lambda$, and after vaccination, the probability of infection is reduced by $(1-\lambda)$ times. Accordingly, a key benefit of vaccination can be quantified as $w_{id}(1-\lambda)$, which means that the more risk reduction, the higher the profit, especially in high-risk settings. When both individuals in the interaction are vaccinated, both parties profit from the superimposed protective effect, reducing the probability of infection by $(1-\lambda^2)$ times, and the corresponding profit is $w_{id}(1-\lambda^2)$. If an individual is unvaccinated and the other has been vaccinated, the unvaccinated individual may receive indirect protection due to the reduced risk of transmission from vaccinated neighbors. The effect of this spillover is modeled by a shared parameter $\xi \in (0,1)$, which measures the extent to which unvaccinated individuals benefit from population protection. An important configuration in this model is that the profit when neither participant is vaccinated is defined as 0. This baseline state means that there are neither vaccination-related costs nor vaccination-related profits, allowing the profit matrix to highlight differences between outcomes based solely on vaccination decisions. At the same time, the matrix covers scenarios where vaccination may produce positive or negative benefits, indicating that under certain conditions, not vaccination may be a more rational choice. Therefore, the vaccination rate will dynamically evolve based on the trade-offs in specific scenarios.

Therefore, we introduce two types of fitness, $\pi_i^0 (t)$ and $\pi_i^1 (t)$, to illustrate the total payoffs of individual $i$ when adopting $X_i (t) = 0$ and $X_i (t)=1$ at time $t$, respectively. This payoff is equal to the sum of the payoffs from the games with all its neighbors. The specific calculation method is given as follows:

\begin{equation}
\left\{\begin{matrix}
\pi_i^0 (t)=p_{nv}\sum_{j\in N_i}X_j (t)+p_{nn}\sum_{j\in N_i} [1-X_j (t)],  \\ 
\pi_i^1 (t)=p_{vv}\sum_{j\in N_i}X_j (t)+p_{vn}\sum_{j\in N_i} [1-X_j (t)].  
\end{matrix}\right.
\end{equation}

Then, we incorporate the following population-level factors into the vaccination game: the number of vaccinated individuals (V), the number of infections caused by the original strain (\(I_1\)) and the mutant strain (\(I_2\)), their respective perceived harm levels, and the mutation rate ($\gamma$) of the original strain. The overall epidemic influence is formally defined as follows:
\begin{equation}
    \mathcal{I}=\frac{\omega_1n_V (t)+w^{I_1}_{Id}n_{I_1} (t)+w^{I_2}_{Id}n_{I_2} (t)+w^{I_2}_{Id}\mu n_{I_1}}{n (t)}, 
\end{equation}
where $n_{Y (t)} (t)$ and $n(t)$ represent the number of individuals in state $Y(t)$ and the total number of individuals at time $t$, respectively. Additionally, $\omega_1$ is the conformity coefficient we configure for the influence of the number of vaccinated people on the strategy selection.

We integrate both the game among neighbors and the overall epidemic situation influence using multiple linear regression to construct the payoff functions for both non-vaccination and vaccination strategies \cite{xiao2019rumor} as follows:
\begin{equation}
    \Pi_i^0=\rho_1\pi_i^0+\rho_2(1-\mathcal{I}),
\end{equation}
\begin{equation}
    \Pi_i^1=\rho_1\pi_i^1+\rho_2\mathcal{I},
\end{equation}
where \(\rho_1\) and \(\rho_2\) represent the weight coefficients of neighbor games and global influence, respectively.

Based on payoff Eqs. (6) and (7) and using the Fermi update rule, we obtain the probability that individual $i$ adopts the vaccination strategy as Eq. (8).
\begin{equation}
P[X_i (t+1)=1|X_i (t)=0]=\frac{1}{1+e^{ (\frac{\Pi_i^0-\Pi_i^1}\kappa)}}. 
\end{equation}
The parameter $\kappa$ characterizes the level of noise in an individual’s strategy update process. 

To present our vaccination strategy model more clearly, the model sketch is shown in Fig. 2.

\subsection{Dynamical State Equations Under the MMCA}
In fact, the selection of vaccination strategies, when reflected in the epidemic dynamic process, corresponds to the state transition from S to V; these two represent the same process viewed from different perspectives.
Then, we can integrate the entire epidemic model, that is, Eqs. (1) and (8), by means of the MMCA \cite{gomez2010discrete, wu2011epidemic}, thus we can get
\begin{equation} 
   \begin{split}
   \theta_i(t) = \frac{1}{1+e^{ (\frac{{\Pi}_i^0-{\Pi}_i^1}\kappa)}}.
   \end{split}
\end{equation}

To capture the heterogeneous transmission rates of nodes, we define the arrival matrix \( R \) to illustrate the probability of node \( j \) contacting node \( i \) based on the degree of node $i$ and the degree weight of node $j$ among neighbors of node $i$. The elements of $R$ are expressed as
\begin{equation}
    r_{ji} = \frac{a_{ji}k_j}{k_i \cdot \sum_{v \in N(i)} k_v},
\end{equation}
where \(a_{ji}\) is the element in the adjacency matrix $A$ of the network, and $N(i)$ represents the neighbors of node $i$.

We then define $q_{1,i} (t)$, $q_{2,i} (t)$, $v_{1,i} (t)$, and $v_{2,i} (t)$ as the probabilities that a susceptible or vaccinated node $i$ is infected by a neighbor carrying strain $I_1$ or $I_2$ at time $t$.

\begin{equation}
\left\{\begin{matrix}
q_{1,i} (t)=1-\displaystyle\prod_{j=1}^{n} [1-r_{ji}\beta_1 P_j^{I_1} (t)],  \\ 
q_{2,i}(t)=1-\displaystyle\prod_{j=1}^{n} [1-r_{ji}\beta_2 P_j^{I_2} (t)],\\
v_{1,i} (t)=1-\displaystyle\prod_{j=1}^{n} [1-r_{ji}\lambda\beta_1 P_j^{I_1} (t)],\\
v_{2,i}(t)=1-\displaystyle\prod_{j=1}^{n} [1-r_{ji}\lambda\beta_2 P_j^{I_2} (t)], \\
\end{matrix}\right.
\end{equation}
where $n$ represents the scale of the network.

Furthermore, we give the formula to describe the probability of node $i$ in state $q_{i} (t)$ and $v_{i} (t)$ being infected by any neighbor at time $t$, respectively:

\begin{equation}
\left\{\begin{matrix}
    q_{i} (t)=1-\displaystyle\prod_{j=1}^{n} [1-r_{ji}\beta_1 P_j^{I_1} (t)]-\displaystyle\prod_{j=1}^{n} [1-r_{ji}\beta_2 P_j^{I_2} (t)],\\
    v_{i} (t)=1-\displaystyle\prod_{j=1}^{n} [1-r_{ji}\lambda\beta_1 P_j^{I_1} (t)]-\displaystyle\prod_{j=1}^{n} [1-r_{ji}\lambda\beta_2 P_j^{I_2} (t)].
    \end{matrix}\right.
\end{equation}

Therefore, we finally present the dynamically evolving equations of the five states as follows: 
\begin{equation}
   \begin{cases}
p^S_i (t+1)= (1-\theta_i(t))(1-q_{i} (t))p_i^S(t)+\alpha p_i^R(t),\\ 
p^{V}_i (t+1)= (p_i^V(t)+\theta_i(t)p_i^S(t))(1-v_{i}(t)),\\
p^{I_1}_i (t+1)=(1-\gamma_1-\mu)p_i^{I_1}(t)+q_{1,i}(t)(1-\theta_i(t))p_i^S(t)\\+v_{1,i}(p_i^V(t)+\theta_i(t)p_i^S(t)),\\
p^{I_2}_i (t+1)=(1-\gamma_2)p_i^{I_2}(t)+ q_{2,i}(t)(1-\theta_i(t))p_i^S(t)+\\v_{2,i}(t)(p_i^V(t)+\theta_i(t)p_i^S(t))+\mu p_i^{I_1}(t),\\
p^{R}_i (t+1)=(1-\alpha)p^{R}_i (t)+\gamma_1p_i^{I_1}(t)+\gamma_2p_i^{I_2}(t), 
\end{cases}
\end{equation}
where $p_i^S (t)+p_i^{V} (t)+p_i^{I_1} (t)+p_i^{I_2} (t)+p_i^R (t)=1.$

\section{Simulation}
In this section, we conduct a series of numerical simulations to systematically explore the influence of various factors on the coupling dynamics between ``vaccination decisions'' and ``dual strain transmission''. Specifically, on the  WS network, we analyze how key parameters such as perceived risk level, strain infection rate, mutation rate, herd effect, initial outbreak size, and vaccine cost and efficacy collectively shape the evolution of different states within the system. 

To explore how vaccination decisions based on evolutionary games affect disease transmission, we define three scenarios with different ``strain perceived risk levels'' based on the profit matrix (Eqs. (2) and (3)) - the perceived risk here corresponds to $w^{I_1}_{Id}$ of the original strain $I_1$ and $w^{I_2}_{Id}$ of the mutant strain $I_2$. Specifically, the criteria for distinguishing low, medium, and high perceived risk groups are the changes from small to large values of the couples of $w^{I_1}_{Id}$ and $w^{I_2}_{Id}$.
\begin{figure*}[ht] 
    \centering
    \vspace*{\fill} 
    \includegraphics[width=\textwidth]{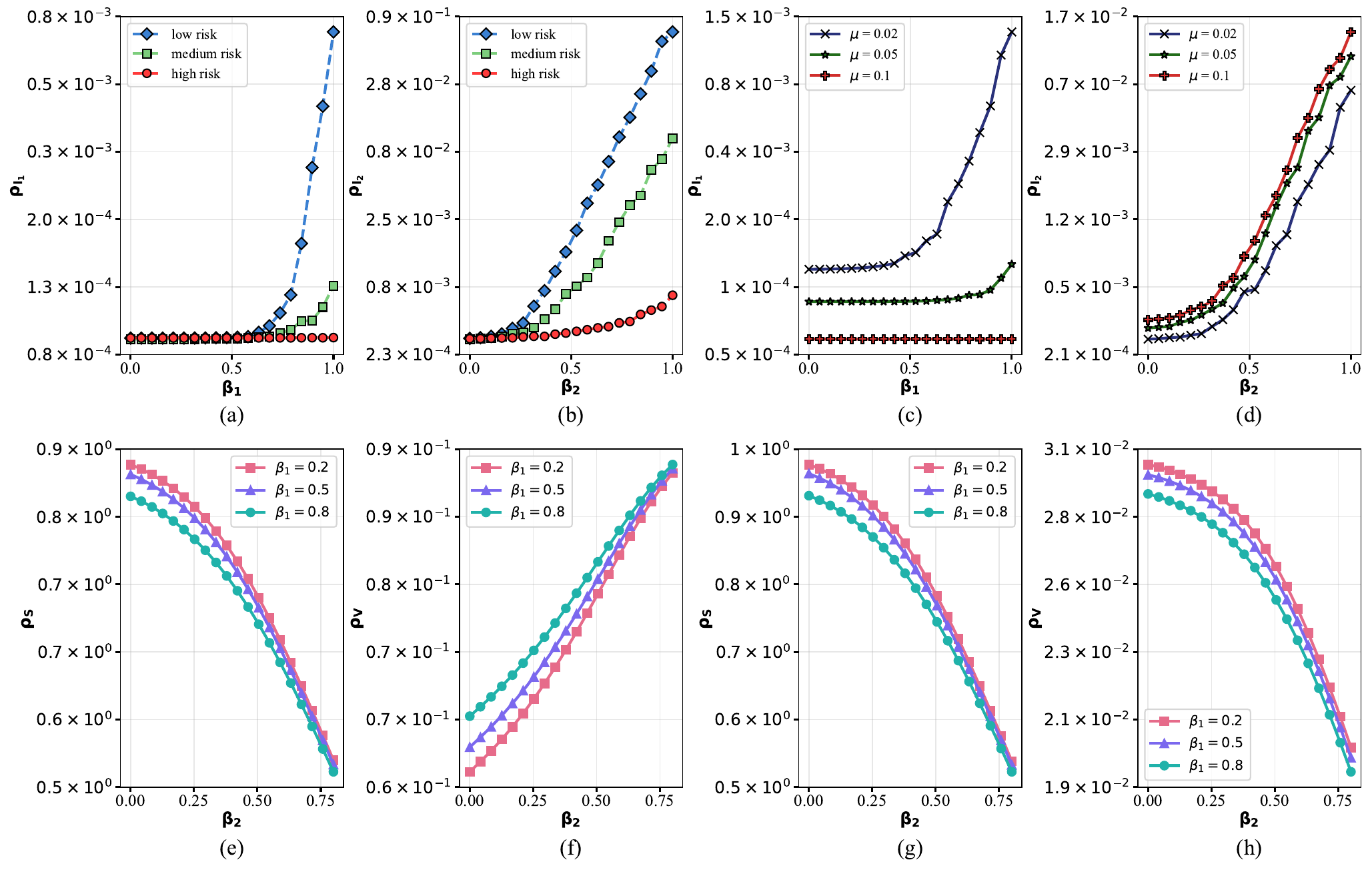}
    \vspace*{\fill} 
    \caption{\textbf{Impact of dual-strain virus infection rates on different perceived risk groups and mutation rates, and their mutual combined effects.} This figure systematically investigates how the infection rates of dual strains (original strain \(\beta_1\), mutant strain \(\beta_2\)), perceived risk levels, and virus mutation rates jointly affect epidemic transmission and vaccination outcomes. For the details, subfigures (a) and (b) show the relationship between the infection rate of the original strain $\beta_1$ and its infection density (represented by the vertical axis), and the relationship between the infection rate of the mutant strain $\beta_2$ and its infection density under different perceived risk scenarios. Subfigures (c) and (d) show the same relationship described above at different mutation rate settings. Subfigures (e) to (h) analyze the combined impact of the infection rate of the two strains on disease transmission. Subfigures (e) and (f) include the impact of the global game on vaccination decisions, and subfigures (g) and (h) represent the scenario without the impact of the global game $W=0$. And subfigures (e) and (g) show the density of individuals S, and subfigures (f) and (h) show the density of individuals V.}
    \label{fig:first_figure}
\end{figure*}

\subsection{Impact of Dual-strain Virus Infection Rates}
In our model, the key parameters that affect the dynamic evolution of the model include the original strain infection rate $\beta_1$, the mutant strain infection rate $\beta_2$, the mutation rate $\mu$, and the fitness functions $\pi_i^0(t)$ and $\pi_i^1(t)$. To delve deeper into the impact of these parameters, Fig. 3 illustrates the overall impact of each study subject on the disease transmission process through the changes in the densities of the three major states in the network: S (susceptible), V (vaccinated), and I (infected, which includes original strain $I_1$ and mutant strain $I_2$). Specifically, in this subsection, we perform repeated simulations of the transmission process by altering parameter settings for different vaccination conditions and mutation rate groups, analyzing the effects of the evolutionary game strategy choices and transmission contexts corresponding to each set of parameters.

As shown in Figs. 3(a)-(d), we find that when the infection rate ($\beta_1$ and $\beta_2$) is relatively low, it does not trigger a large-scale disease outbreak; however, once $\beta_1$ and $\beta_2$ exceeds a certain critical threshold, for example, $\beta_1 = 0.6$ for the high-risk group in Fig. 3(a), the disease transmission continues, and the number of infected individuals increases with increasing $\beta_1$ and $\beta_2$, thereby expanding the population coverage of the epidemic. Notably, the simulation results indicate that the infection density of the original strain remains at a very low level (on the order of $10^{-4}$, whereas the infection density of the mutant strain is significantly higher (on the order of $\geqslant 10^{-2}$), which is clearly due to the higher infection rate of the mutant strain. In addition, when comparing Figs. 3(a) and 3(b), we find significant differences in the state dynamics among groups with different perceived risk levels: the higher the perceived risk of a group, the stronger its effect in suppressing disease transmission, and the lower the final infection density. In the high perceived risk group, the infection density remains nearly zero, suggesting that a high level of individual vigilance toward the disease and extensive vaccination behavior under high-risk conditions prevent widespread transmission on the network. In addition, from Figs. 3(c) and 3(d), it can be observed that the mutation rate $\mu$ also has a significant effect on the infection distribution of the two strains. When the infection rate of the original strain $\beta_1$ is the same, the higher the $\mu$, the more original strain-infected individuals mutate into mutant strain-infected individuals, leading to a decrease in the density of $I_1$ and an increase in the density of $I_2$. Moreover, as $\beta_1$ increases, the density of the original strain-infected individuals increases, allowing more individuals to mutate into mutant strain-infected individuals, so that this effect becomes increasingly pronounced.
\begin{figure*}[h!]
    \centering
    \vspace*{\fill} 
    \includegraphics[width=\textwidth]{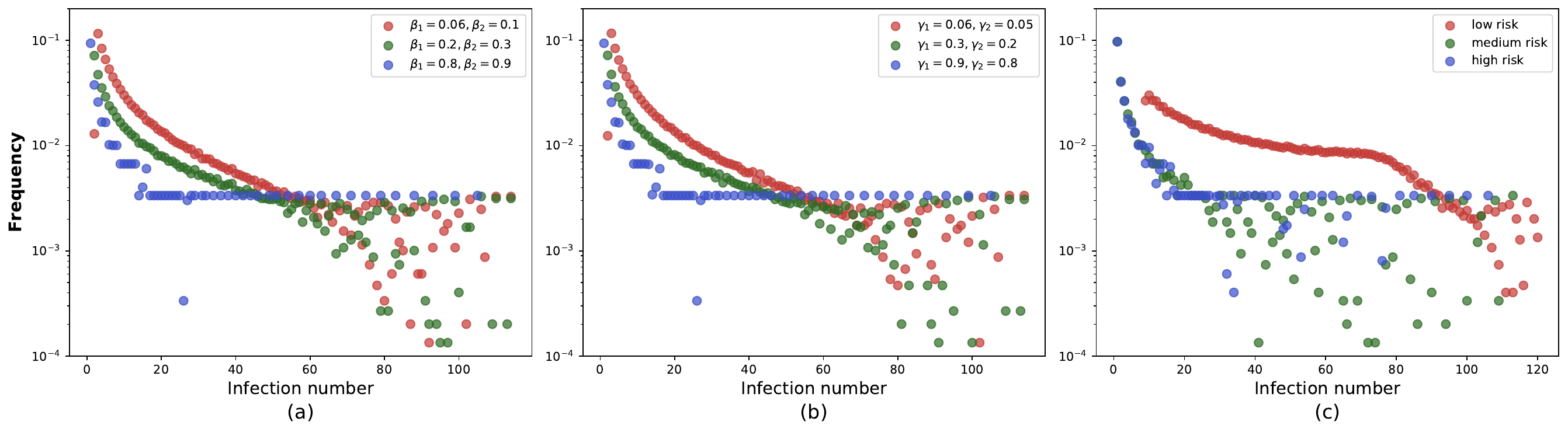}
    \vspace*{\fill}
    \caption{\textbf{Frequency of infected population.} This figure presents simulation results regarding the occurrence frequency of different infected population numbers, modulated by three key parameters: infection rate, recovery rate, and perceived risk (each parameter is categorized into low, medium, and high levels). Subfigures (a), (b), and (c) respectively correspond to the three parameter groups ($\beta_1,\beta_2$, $\gamma_1,\gamma_2$, and perceived risk). In each subfigure, the red, green, and blue lines correspond to the low, medium, and high levels within each group, respectively. The horizontal axis represents the number of infected individuals observed throughout all iterations, and the vertical axis indicates the frequency of occurrence for each infected count during the simulation. }
    \label{fig:second_figure}
\end{figure*}

Furthermore, we analyze the combined effects of the original and mutant strains: we first consider the population changes in S individual as shown in Figs. 3(e) and (g), and we can find that the increase in the infection rate of either strain will prompt more S individuals to choose to be vaccinated, which in turn will lead to a decrease in the density of S individuals $\rho_S$, and Fig. 3(g) shows a similar trend for the corresponding simulation scenario (where inoculation decisions are not affected by the global game). Then, we observe the population changes of V individual $\rho_V$, comparing Figs. 3(f) and (h), and we find that the change pattern is significantly different: in Fig. 3(f), an increase in the infection rate of any strain will increase the number of vaccinated individuals; In contrast, in a scenario without global game impact (Fig. 3(h)), individuals do not incorporate population-level outbreak information into vaccination decisions, so changes in infection rates do not directly affect vaccination behavior. Conversely, as the number of infected individuals increases, the density of S individuals and V individuals decreases over time.

\subsection{The Frequency of the Infection Number in an Epidemic}
In real epidemic transmission scenarios, changes in the number of infections are the central concern of all agencies. In this subsection, we study the changes in the number of infections among groups with different perceived risks within the same network throughout the transmission process. We illustrate these changes through the frequency distribution corresponding to specific numbers of infections. We set up comparative experiments at three levels---low, medium, and high---for perceived risk, infection rate, and recovery rate, as shown in Fig. 4.

Fig. 4(a) shows that the lower the infection rate, the better the suppression effect on epidemic spread. In addition, we note that the number of infected individuals mainly remains at 6\% of the total population and never exceeds 10\%. This phenomenon is more pronounced in groups with lower infection rates, where the frequency of low infection numbers is higher. This further indicates that reducing the transmission rate will significantly slow down the speed and extent of epidemic spread. Moreover, we find that the trend presented in Fig. 4(b) is consistent with Fig. 4(a). However, it is worth noting that groups with lower recovery rates have a lower average number of infected individuals. We argue this is because under low recovery rate conditions, the infection density experiences a temporary surge, which increases the overall influence of vaccination and promotes the vaccination game mechanism. The behavioral response generated thereby strengthens the control over transmission, offsetting the impact of reduced natural immunity, ultimately keeping the number of infections in a lower range. This also reflects that the vaccination behavior based on evolutionary game theory in our model improves the traditional epidemic transmission process and provides a more accurate fit to real-world scenarios. Finally, Fig. 4(c) shows that groups with low-risk perception exhibit high infection frequencies across all infection number ranges, including a large-scale infection scenario involving up to 120 people, which is not observed in medium- and high-risk perception groups. This suggests that a low-risk perception can compromise the effectiveness of epidemic control across the entire network. In contrast, the results for medium- and high-risk perception groups are almost identical, indicating that once risk perception reaches a certain threshold, it can effectively suppress disease transmission.

\begin{figure*}[h!]
    \centering
    \vspace*{\fill} 
    \includegraphics[width=\textwidth]{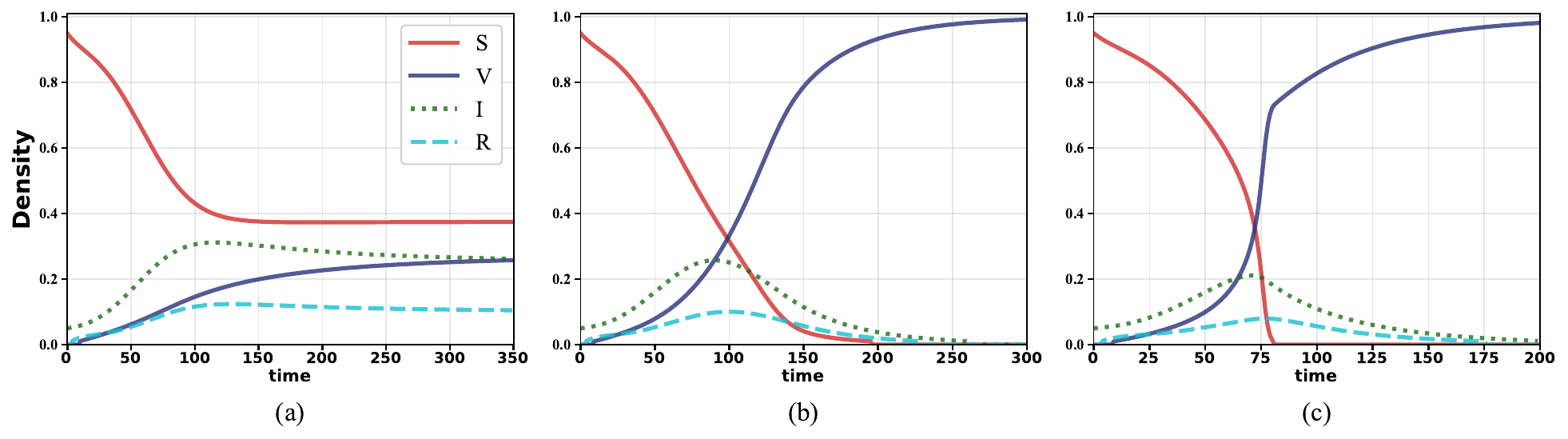}
    \vspace*{\fill}
    \caption{\textbf{Temporal evolution of epidemic spread and strategy selection under different herding coefficients.} This figure characterizes epidemic dynamics by three core metrics, the number of infected individuals, the densities of S and V individuals, and further reveals how herding behavior modulates the temporal evolution of epidemic spread and individual vaccination strategy selection. The epidemic dynamics are characterized by the number of infected individuals and the densities of S and V individuals. In subfigures (a)–(c), the green dotted, red solid, blue solid, and cyan dashed lines represent the time-varying densities of I,  S, V, and R individuals, respectively. Among them, subfigures (a)–(c) correspond to three scenarios of low, medium, and high conformity tendencies, with the conformity coefficient \(\omega_1\) being 0.1, 0.6, and 0.9, respectively.  }
    \label{fig:second_figure}
\end{figure*}

\begin{figure*}[h!]
    \centering
    \vspace*{\fill} 
    \includegraphics[width=\textwidth]{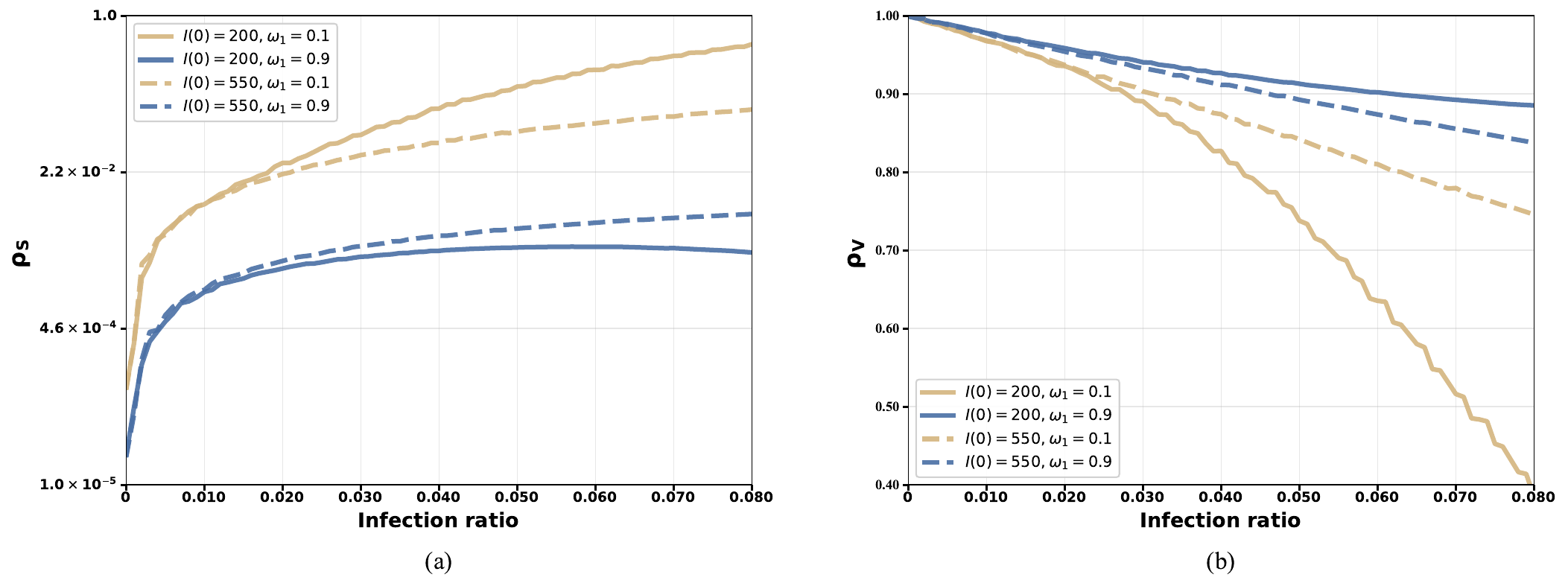}
    \vspace*{\fill}
    \caption{\textbf{Evolution analysis of vaccination status and infection density under four scenarios.} This figure explores the evolutionary co-influence between the conformity phenomenon and initial infection density across four combined scenarios, which are defined by two levels of conformity coefficient (\(\omega_1\)) and two sizes of initial infected population. In both subfigures, the yellow and blue lines correspond to scenarios with conformity coefficients $\omega_1$ of 0.1 and 0.9, respectively. The dashed and solid lines represent initial infected population sizes of 550 and 200, respectively. Specifically, subfigure (a) illustrates the density of individual S as a function of infection density, while subfigure (b) depicts the corresponding density of individual V under varying infection densities. }
    \label{fig:second_figure}
\end{figure*}

\subsection{The Impact of Herd Behavior Coefficient and Outbreak Scale on Epidemic}
In our model, the evolutionary game mechanism in vaccination behavior is an important innovation and also the core mechanism. This process simultaneously affects the densities of the three key states S, V, and I (including $I_1$ and $I_2$). Furthermore, the conformity behavior incorporated into the game framework is a key factor of interest in this study. Therefore, this subsection simulates the evolution of the three states S, V, and I over time based on their density, evaluating the scenarios under different initial epidemic sizes and various conformity coefficients $\omega_1$. The relevant results are shown in Figs. 5 and 6.
\begin{figure*}[h]
    \centering
    \vspace*{\fill} 
    \includegraphics[width=\textwidth]{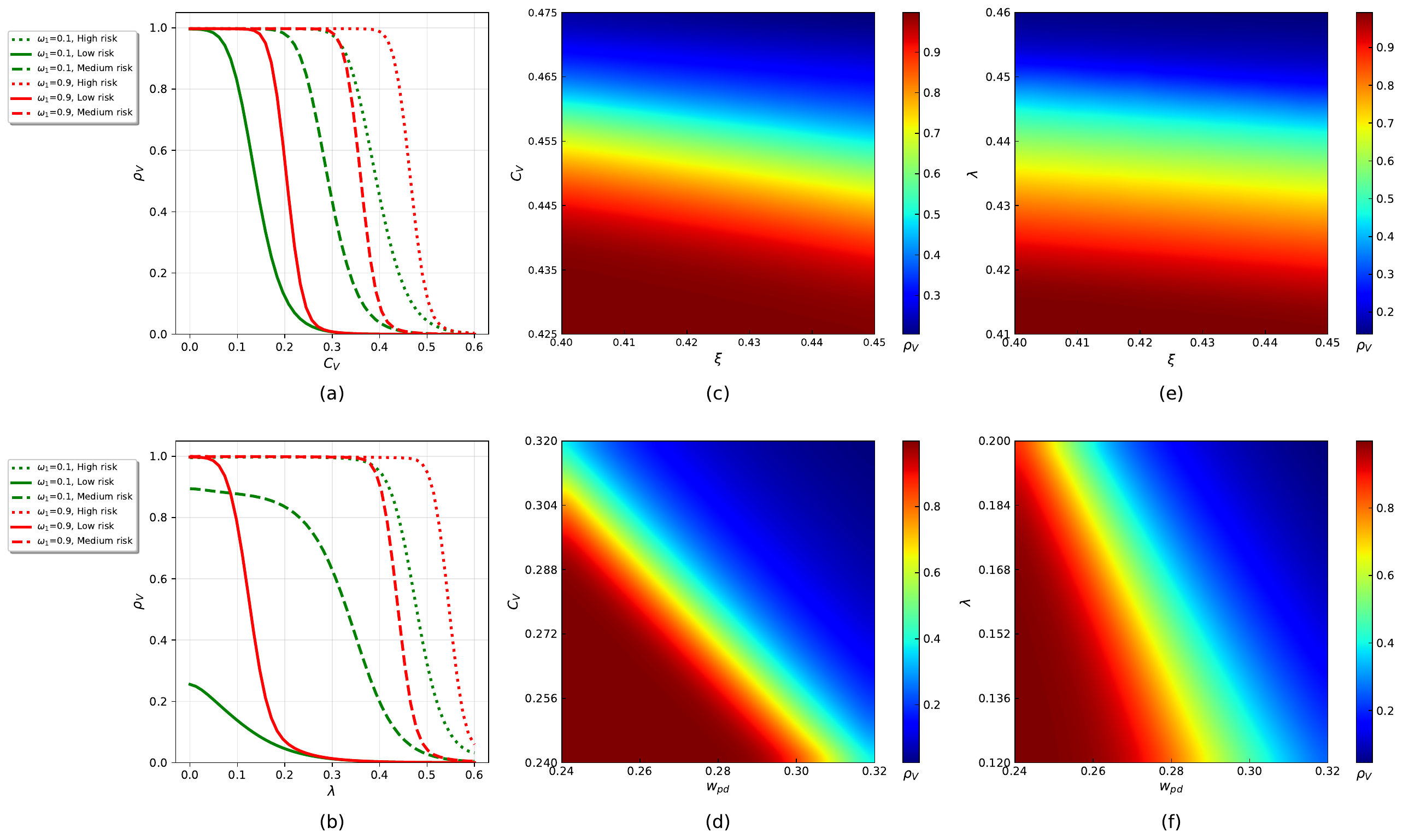}
    \vspace*{\fill}
    \caption{\textbf{Impact of vaccination-related parameters.} This figure further evaluates how six core vaccination-related parameters---vaccination cost (\(C_V\)), vaccine efficacy (\(\lambda\)), shared parameter (\(\xi\)), potential negative benefit (\(w_{pd}\)), perceived risk, and conformity coefficient (\(\omega_1\))---jointly affect population-level vaccination density.  Subfigures (a) and (b) present line graphs depicting how vaccination density varies with increasing vaccination cost (\(C_V\)) and vaccine efficacy (\(\lambda\)), respectively. Green lines indicate scenarios with low conformity (\(\omega_1=0.1\)), and red lines represent high conformity (\(\omega_1=0.9\)). Solid, dashed, and dash-dotted lines correspond to low, medium, and high perceived risk groups, respectively. Subfigures (c) and (e) show heatmaps illustrating the joint effects of the sharing parameter combined with vaccination cost and vaccine efficacy, respectively, on vaccination density. Subfigures (d) and (f) present heatmaps examining the combined influence of potential negative benefits of vaccination, together with vaccination cost and vaccine efficacy, on vaccination density. In all heatmaps, color gradients from blue to red indicate vaccination density values from low to high. }
    \label{fig:second_figure}
\end{figure*}
By first comparing Figs. 5(b) and 5(c), we can observe that although both scenarios ultimately achieve disease elimination, the process of disease elimination under a high conformity coefficient is significantly faster. At the same time, under high conformity, the number of individuals in the S group decreases more noticeably, and the number of individuals in the V group correspondingly increases more noticeably. The slopes of the density curves for S and V are steeper, indicating that their numbers change more intensely. We suggest that this highlights the significant impact of collective behavior on vaccination speed and effectiveness. Furthermore, comparing Fig. 5(a) with Figs. 5(b)–5(c), it is evident that when the conformity coefficient is extremely low, meaning that individuals’ decisions to get vaccinated rely mainly on the games among neighbors and the pressure of infection density in the network, disease elimination cannot be achieved; instead, the system enters a local equilibrium state, with the infection density remaining at a continuously low level. This result emphasizes the importance of moderate social influence in promoting effective epidemic control and further suggests that public health strategies that reasonably encourage collective protective behavior may enhance the suppression of disease transmission.

In addition, as shown in Fig. 6, the initial number of infections and the herd coefficient $\omega_1$ together affect vaccination behavior. Comparing the dotted line with the solid line, it can be seen that under the low conformity coefficient, when the infection density is the same, the more the number of initial infections, the more S individuals choose to be vaccinated. In contrast, at high conformity coefficients, this relationship is reversed: the lower the initial number of infections, the higher the vaccination rate of S individuals at the same infection density.
We argue this reversal is because higher initial infection densities encourage early vaccination of large populations, which can quickly contain disease spread and eliminate disease, while the vaccinated population is still relatively small. In this case, high conformity will weaken the role of evolutionary game interaction between neighbors in subsequent decision-making, resulting in vaccination levels stabilizing at a low equilibrium. Additionally, we further compare the blue line and the yellow line, that is, the groups with different conformity coefficients, which shows that under similar infection densities, the number of vaccinated people is significantly higher and the number of susceptible people is significantly lower in the group with a higher conformity coefficient, which shows the importance of social imitation behavior in vaccination decision-making. However, this effect weakens in the case of a larger initial outbreak. We note that this is because higher initial infection levels and fast-spreading infections can lead to early herd infection in some vulnerable populations who are infected before vaccination is complete, thereby weakening the herd effect. Finally, in Fig. 6(b), the gap between the susceptible density of the low-conformity group and the high-conformity group gradually widened as the infection density decreased, which further confirmed the above conclusion.

\subsection{Impact of Vaccination-Related Parameters}
In addition to the fact that infections from the two major pathogen strains play a key role in disease transmission, the vaccination dynamics in our model are also intrinsically linked to the overall trajectory of the epidemic and have a significant impact on transmission dynamics. In this subsection, based on the previous simulation experiments, we further assess the impact of key vaccine-related parameters on vaccination coverage, with the related results shown in Fig. 7.

Overall, Figs. 7(a) and 7(b) indicate that there are significant thresholds for vaccination density in relation to vaccination cost and vaccine efficacy. We find that in groups with low-risk perception and high conformity coefficients, almost no one would get vaccinated when the cost coefficient exceeded 0.3; when it was below this value, vaccination rates rose sharply, eventually approaching full coverage. Vaccine efficacy also exhibits a similar threshold, around 0.35. Similar thresholds are present in other groups as well, suggesting that in real life, thresholds could be estimated based on various factors to formulate threshold-based vaccine cost control strategies. Meanwhile, comparing trends across different risk perception groups shows that under the same cost and conformity effects, the density of V individuals decreases as risk perception goes from high to low. It is noteworthy that even under adverse conditions for promoting vaccination, such as higher vaccine costs or lower conformity effects, groups with high-risk perception can still maintain relatively high vaccination density, further reflecting the phenomenon of large-scale vaccination driven by collective concern due to widespread epidemic transmission. In addition to the influence of costs and perceived risks, this simulation further confirms that conformity also plays an important role in vaccination behavior: in all scenarios, groups with higher conformity coefficients had higher vaccination rates, and they were able to maintain relatively high vaccination rates even under unfavorable conditions.

Now, we will verify the effects of various parameters through the parameter heatmap experiments shown in Figs. 7(c)–(f). First, in Fig. 7(c), an increase in the sharing parameter (i.e., the indirect profit gained by unvaccinated individuals from others’ vaccination) slightly reduces the vaccination density. Similar effects were observed in Fig. 7(e) when examining the combined influence of the sharing parameter and vaccine efficacy. Additionally, we find that these two subplots exhibit a distinct vertical stratification phenomenon, further indicating that the higher the vaccine cost and the lower its efficacy, the more significantly the vaccination rate declines. We also note that Figs. 7(d) and 7(f) display clear diagonal symmetry, suggesting that perceived negative effects of vaccination (such as side effects) significantly influence vaccination behavior. The more severe the perceived negative impact, the fewer people get vaccinated, and the overall vaccination density decreases. Comparing Figs. 7(d) and 7(f), we can see that the gradient symmetry in Fig. 7(d) is more pronounced, indicating that vaccine cost has a stronger effect on vaccination rates than vaccine efficacy. We believe that these comparative results successfully demonstrate that high-risk perception and proactive preventive awareness can effectively curb the rapid, large-scale spread of the virus. Moreover, vaccine cost and efficacy greatly influence people’s willingness to get vaccinated. These research conclusions guide for future studies, and relevant institutions or individuals involved in vaccine development and management should pay more attention to the importance of these influencing factors.

\section{Conclusion and Outlook}
We combine evolutionary game theory with population-level epidemic spread to establish a novel dynamic model that couples propagation dynamics with evolutionary game behavior. At the transmission level, we propose an extended SIRS model that simultaneously incorporates vaccination behavior and the spread of mutant strains; at the vaccination game-theoretic level, we consider both neighbor-based multi-factor game strategies and the effects of the overall epidemic situation and herd behavior. Finally, by coupling strategic decision-making with epidemic spread, we derive the corresponding MMCA state transition probabilities. 
Through multi-angle simulations under different parameter settings, the advantages of our model are validated. Different levels of perceived disease risk significantly influence transmission dynamics; the strength of herd mentality strongly affects vaccination willingness, thereby altering the process and outcome of disease spread; in addition, we point out that considering a payoff structure adaptable to the population state allows strategies to be dynamically adjusted according to epidemiological conditions. These features collectively enhance the ability of relevant authorities to resist, control, and ultimately eliminate outbreaks. On this basis, we also conduct multi-scenario simulation experiments to confirm the model's effective containment capabilities: within our model framework, diseases with moderate to low infection and mutation rates cannot sustain transmission; even in the face of large-scale outbreaks, the model can promote rapid epidemic control in a short period. These results validate the model’s rationality and practical applicability. Finally, sensitivity analysis of vaccine-related parameters further highlights the specificity and practicality of the model and reveals the impact of various detailed mechanisms. These findings provide actionable guidance for policymakers: interventions that reasonably reduce vaccine costs, improve vaccine efficacy, and mitigate adverse vaccine reactions are expected to increase vaccination rates and strengthen epidemic containment. The simulation results also offer practical guidance for public health management and personal preventive strategies, such as when to strengthen protective measures, how to enhance risk awareness, and whether to prioritize controlling viral mutations or increasing treatment rates.  

However, several aspects still warrant further research. For example, how the herd effect interacts with misinformation or rumors, whether risk perception can be dynamically adjusted through infection probability, and how the model performs in larger-scale adaptive networks. In addition, our closed-network approach does not account for demographic dynamics such as birth, death, and migration, and incorporating these processes into the model may represent an important direction for future research.

\section*{Acknowledgements}
This work was supported in part by the Natural Science Foundation of Chongqing under Grant NO. CSTB2025YITP-QCRCX0007; in part by the National Natural Science Foundation of China (NSFC) under Grant NO. 62206230.







\end{document}